\documentclass[aps,twocolumn,superscriptaddress,prb,showpacs,nofootinbib]{revtex4-1}

\usepackage{graphicx} 
\usepackage{amsmath} 
\usepackage{epstopdf}
\usepackage[abs]{overpic} 
\usepackage{subfigure} 
\usepackage{cancel}
\usepackage{nicefrac}

\usepackage{comment}

\begin{document}
	
\title{Theory of the optical spinpolarization loop of the nitrogen-vacancy center in diamond}
	
	\author{Gerg\H{o} Thiering} \affiliation{Wigner Research Centre for Physics,
		Hungarian Academy of Sciences, PO Box 49, H-1525, Budapest, Hungary}
	\affiliation{Department of Atomic Physics, Budapest University of Technology and
		Economics, Budafoki \'ut 8., H-1111 Budapest, Hungary}
	
	\author{Adam Gali} \email{gali.adam@wigner.mta.hu} \affiliation{Wigner Research
		Centre for Physics, Hungarian Academy of Sciences, PO Box 49, H-1525, Budapest,
		Hungary} \affiliation{Department of Atomic Physics, Budapest University of
		Technology and Economics, Budafoki \'ut 8., H-1111 Budapest, Hungary}

\begin{abstract}
Nitrogen-vacancy (NV) center in diamond is of high importance in quantum information processing applications. The operation of NV center relies on the efficient optical polarization of its electron spin. However, the full optical spinpolarization process, that involves the intersystem crossing between the shelving singlet state and the ground state triplet, is not understood. Here we develop a detailed theory on this process which involves strong electron-phonon couplings and correlation of electronic states that can be described as a combination of pseudo and dynamic Jahn-Teller interactions together with spin-orbit interaction. Our theory provides an explanation for the asymmetry between the observed emission and absorption spectra of the singlet states. We apply density functional theory to calculate the intersystem crossing rates and the optical spectra of the singlets and we obtain good agreement with the experimental data. As NV center serves as a template for other solid-state-defect quantum bit systems, our theory provides a toolkit to study them that might help optimize their quantum bit operation.
\end{abstract}
\maketitle

\section{Introduction\label{sec:Intro}}

\begin{figure*}
	\includegraphics[width=\textwidth]{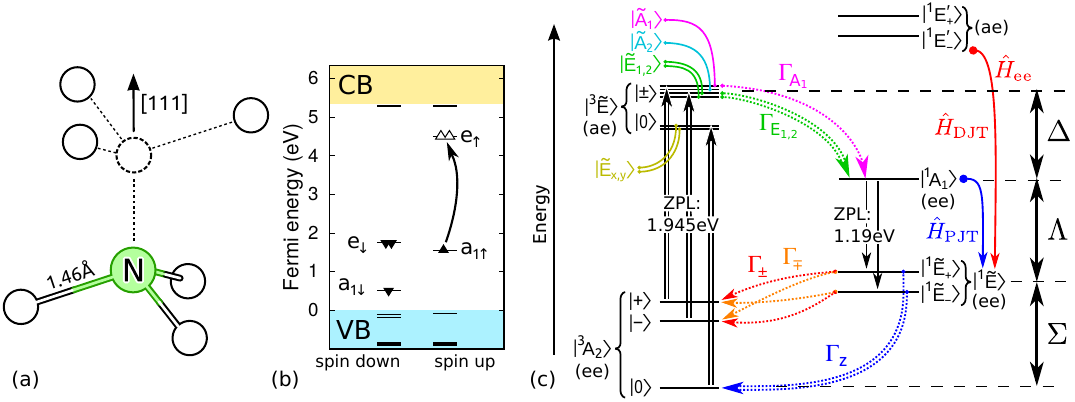}
	\caption{\label{fig:NV} Properties of NV center in diamond. (a) Geometry structure showing the symmetry axis of $C_{3v}$ symmetry. The vacancy is the dotted circle whereas the solid circles depict carbon atoms in the diamond lattice. (b) Calculated HSE06 Kohn-Sham levels in the diamond band gap between the valence (VB) and conduction (CB) band.  (c) Many-electron states are expressed in two-particle Slater-determinants in the parentheses (see~Eqs. \eqref{eq:State1}-\eqref{eq:State6}). The many-electron levels are also depicted with the measured zero-phonon-lines (ZPL). The zero-field-splittings in the triplet manifolds are artificially scaled up by five orders of magnitude for the sake of clarity. We label the possible intersystem crossing rates ($\Gamma$s) with colored dotted arrows that participate in the spin polarization cycle. We label the radiative transitions in the aforementioned cycle with solid black arrows. We refer to the vibronic states (coupled electron-phonon wavefunctions) by tilde labels above the many-electron states. The nature of the vibronic $^1\widetilde{E}$ state is explained (blue and red solid curves and texts) where $ee$, PJT and DJT stands for electron-electron, pseudo Jahn-Teller, and dynamic Jahn-Teller interactions, respectively.}	
\end{figure*}

The most known point defect in diamond is the nitrogen-vacancy (NV) center\cite{duPreez:1965} which acts as a quantum bit for solid state quantum information processing applications \cite{Gruber1997, Jelezko2004, Ladd2010, Awschalom2013, doherty2013nitrogen}. NV center is a negatively charged complex which consists of a substitutional nitrogen next to a vacancy in diamond [see Fig.~\ref{fig:NV}(a)]. The defect has an $S=1$ ground state with milliseconds coherence time at room temperature in $^{12}$C enriched diamond samples \cite{Balasubramanian2009}, and can be optically excited in the visible~\cite{duPreez:1965}. Under illumination, the electron spin is preferentially populated at the $m_s=0$ spin state over the $m_s=\pm1$ states \cite{Nizovtsev2003, Harrison2004, Manson2006, Robledo2010, Robledo2011}. The robust spin-selective fluorescence \cite{Gruber1997} and photocurrent \cite{Bourgeois2015} are the most important features of this center that can be used for quantum bit initialization and readout schemes. 

Group theory considerations~\cite{Lenef1996, Maze2011, Doherty2011} together with luminescence~\cite{Rogers2008} and absorption~\cite{Kehayias2013} measurements imply that two singlet levels, $^1A_1$ and $^1E$, that are separated by 1.19~eV, reside between the $^3A_2$ ground state and $^3E$ excited state triplet levels. In the optical spinpolarization cycle both singlet states play a role [see Fig.~\ref{fig:NV}(c)]. In the upper branch, highly spin-selective intersystem crossing (ISC) occurs between the $^3E$ triplet and $^1A_1$ singlet caused by the phonon-mediated spin-orbit interaction. Combined photoluminescence excitation (PLE) measurements~\cite{Goldman2015a, Goldman2015b} and perturbation theory on the ISC rates analyzed this process in detail. We note here that the observed multiple rates between the $^3E$ triplet substates and the singlet $^1A_1$ goes against the selection rules~\cite{Maze2011, Doherty2011} that would allow only a single scattering channel [$\Gamma_{A_1}$, i.e., purple dotted arrow in Fig.~\ref{fig:NV}(c)]. The multiple rates can be naturally explained by invoking the dynamic Jahn-Teller effect on the $^3E$ state which can account well to the ratio of the observed ISC rates at cryogenic temperatures~\cite{Thiering2017}. The dynamic Jahn-Teller effect is a special description of a strongly coupled electron-phonon system that mixes the pure electronic substates of $^3E$ with each other that results in vibronic states that are labeled by tilde in Fig.~\ref{fig:NV}(c). The mixture of $A_1$ into $\widetilde{E}_{1,2}$ by phonons results in ISC from $\widetilde{E}_{1,2}$ toward $^1A_1$ [$\Gamma_{E_{1,2}}$, i.e., green dotted arrow in Fig.~\ref{fig:NV}(c)]. 

However, the ISC process in the lower branch, i.e., between $^1E$ and $^3A_2$, is still not understood. By considering single determinant $^1E$ state built up from the $e$ orbitals [see Fig.~\ref{fig:NV}(b) and Refs.~\onlinecite{Maze2011, Doherty2011}], group theory indicates that no ISC is allowed between this singlet and the $^3A_2$ triplet. On the other hand, the measured lifetime of the $^{1}E$ state is $T_{E}=371 \pm 6$~ns at cryogenic temperatures~\cite{Robledo2011} and ISC toward the $m_s=0$ of $^3A_2$ [$\Gamma_z$, blue dotted arrow in Fig.~\ref{fig:NV}(c)] should be effective to observe spinpolarization in NV center. The measured $T_{E}$ is temperature dependent and decreases down to $\sim$165~ns at room temperature~\cite{Robledo2011}. The temperature dependence could be well understood by a stimulated phonon emission process with an energy of $16.6\pm0.9$~meV~\cite{Robledo2011}. By means of spin control and pulsed optical excitation of NV center, the spin-dependent ISC rates of $^1E$ were extracted at room temperature~\cite{Robledo2011}. Interestingly, it was deduced that the ISC rates from $^1E$ to $m_s=0$ and $m_s=\pm1$ [$\Gamma_\pm$ and $\Gamma_\mp$, red and orange dotted arrows in Fig.~\ref{fig:NV}(c), respectively] are comparable. As the ISC is dominantly spin-selective in the upper branch, this conclusion is not contradictory to the measured $>$90\% optical spinpolarization in the triplet ground state. From these experimental data we may conclude that $^1E$ is linked to the $m_s=\pm1$ in the $^3A_2$ state by spin-flipping transitions (rates of $\Gamma_\pm$ and $\Gamma_\mp$) and to the $m_s=0$ by the rate of $\Gamma_z$, where $T_{E}^{-1}=\Gamma_{z}+\Gamma_{\pm}+\Gamma_{\mp}$=2.70~MHz at cryogenic temperatures. In a previous measurement~\cite{Acosta2010}, a similar value, $T_E^{-1}$=2.16~MHz was deduced. We emphasize that understanding the mechanisms governing the ISC between $^1E$ and $^3A_2$ is very important as this ISC is responsible for closing the optical spinpolarization loop of the NV center which is the base of quantum bit initialization and readout. 

In this study we derive the electron-phonon assisted spin-orbit interaction between the $^1E$ state and the $^3A_2$ state. We show that the nature of $|^{1}E\rangle$ is highly complex as it involves significant electron-phonon coupling to the $|^1A_1\rangle$ and electron-electron interaction to the $|^1E^\prime\rangle$ manifold. The former interaction can be described in the frame of pseudo Jahn-Teller (PJT) effect~\cite{bersuker2012vibronic, bersuker2013jahn} and is responsible for closing the optical spinpolarization loop of NV center. The latter one brings a dynamic Jahn-Teller (DJT) character to the $|^{1}E\rangle$, and explains the observed ISC rate toward the $m_S=\pm1$ states of $^3A_2$ ground state. We identify a novel interplay between PJT and DJT interactions that determine the phonon sideband of the photoluminescence (PL) spectrum of the singlets~\cite{Rogers2008}. Our results also explain the appearance of a new feature in the PL phonon sideband upon the applied uniaxial stress~\cite{manson2010optically} and the asymmetry in the phonon sidebands of the PL and absorption~\cite{Kehayias2013} spectra of the singlets. We use \emph{ab initio} wavefunctions and adiabatic potential energy surfaces (APES) to quantify the strength of interactions and the corresponding temperature dependent ISC rates where the latter ones show good agreement with the experimental data.  

The paper is organized as follows. Sec.~\ref{sec:2} describes the electronic structure of NV center and establishes the nomenclature of the paper. Then we describe the \emph{ab initio} methods in Sec.~\ref{sec:3}. We present the theory of psedo Jahn-Teller effect and the dynamic Jahn-Teller effect brought by electron-electron correlation on the shelving singlet state in Sec.~\ref{sec:4} which contains the main idea of the paper. Sec.~\ref{sec:5} contains the main results of the paper, where we apply \emph{ab initio} calculations to calculate the optical spectra and ISC rates based on the developed theories in Sec.~\ref{sec:4}. Finally, we summarize and conclude our results in Sec.~\ref{sec:6}.

\section{Methodology on atomistic simulations \label{sec:2}}

We apply \emph{ab initio} wavefunctions and APES for determining the electron-phonon couplings, calculating the optical spectra and ISC rates in the framework of  
spin-polarized density functional theory (DFT) as implemented in the \textsc{vasp} 5.4.1 code~\cite{Kresse:PRB1996}. We use the HSE06 hybrid functional~\cite{Heyd03, Krukau06} within DFT that technique reproduces the experimental band gap and the charge transition levels in Group-IV semiconductors within 0.1~eV accuracy \cite{Deak:PRB2010}. We converged the electronic structure with self-consistent cycles on Kohn-Sham orbitals with a low energy cutoff (370~eV)  within the applied projector-augmentation-wave-method
(PAW)~\cite{Blochl:PRB1994, Blochl:PRB2000}. The total energies of the excited states were calculated within the $\Delta$SCF method~\cite{Gali:PRB2009} that provides accurate zero-phonon-line (ZPL) energy and Stokes-shift for the optical excitation spectra of the triplets of NV center.

The negatively charged NV defect is modeled in a 512-atom supercell and $\Gamma$-point is applied to sample the Brillouin-zone. We determine the equilibrium position of ions by minimizing the quantum mechanical forces acting on them below the threshold of 10$^{-2}$~eV/\AA . In the APES, the $C_{1h}$ distorted geometries exhibit the deepest energy configurations. The APES around high $C_{3v}$ symmetry configurations toward the low $C_{1h}$ symmetry configurations is calculated. The corresponding normal modes of the $E$-symmetry phonons participating in the distortion are calculated in the $^3A_2$ ground state by using the quasi-harmonic approximation and finite difference method on the quantum mechanical forces.

In the calculation of ISC, the spin-orbit coupling should be determined between the corresponding states. In our previous work, we determined the $z$-component of the spin-orbit coupling ($\lambda_z$) accurately by our DFT method~\cite{Thiering2017} that resulted in $\lambda_z$=15.78~GHz. We also found that the calculation of the perpendicular component of the spin-orbit coupling ($\lambda_\perp$) requires approximations (see Ref.~\onlinecite{Thiering2017} for discussion) that lead to a significant overestimation in $\lambda_\perp$. Therefore, we use $\lambda_\perp$ here as a parameter which should be greater than the calculated $\lambda_z$.  

The phonon sideband in the absorption spectrum is described by Huang-Rhys (HR) theory\cite{Huang1950} that was previously implemented for DFT supercell methodology~\cite{Alkauskas2014}. The contribution of the $a_1$ and $e$ phonons in the phonon sideband of the PL spectrum for the triplets of NV center was obtained by this methodology~\cite{Thiering2017} where the latter is responsible for the DJT effect in the $^3E$ excited state. 

Finally, we note that our HSE06 DFT method cannot directly calculate the many-body $^1A_1$ and $^1E$ singlet states. Therefore, the energy gap between the $^1E$ and $^3A_2$ states ($\Sigma$) is a parameter. The full \emph{ab initio} description requires to go beyond Kohn-Sham DFT that can describe strong correlation between localized electrons~\cite{Delaney2010, Ranjbar2011, Choi2012}. On the other hand, we will show that HSE06 DFT singlet states within the $\Delta$SCF framework can be employed to derive the parameters for the Jahn-Teller Hamiltonians and estimate the strength of correlation between the $^1E^\prime$ and $^1E$ states.

\section{Preliminaries}\label{sec:3}

Here we define the basic nomenclature of the paper. We note that the orbitals and levels of NV center from DFT calculations have been already published in several papers~\cite{Goss1996, uszczek2004, Larsson2008, Lin2008, Gali2008, Delaney2010, Ma2010, Ranjbar2011, Choi2012}. Furthermore, the corresponding many-body states and the spin-orbit couplings between them were also thoroughly analyzed~\cite{Maze2011, Doherty2011}. Instead of frequently referring to these papers, we rather write explicitly the corresponding wavefunctions and interactions that we use in the entire paper.   

\subsection{Electronic structure}

NV defect introduces an $a_1$ and a double degenerate $e$ level in the gap [$a$ and $e$ orbitals in Fig.~\ref{fig:NV}(b)] that are occupied by four electrons in the relevant negatively charged state. In the hole picture, two holes are left on the $e$ orbital in ground state electron occupation that we simply label as $(ee)$. The many-body ground state triplet state with labeling the $m_S=\{+, 0, -\}$ spin projections can be described as
\begin{equation}
\left.\begin{array}{c}
\left|^{3}A_{2}^{+}\right\rangle \\
\left|^{3}A_{2}^{0}\right\rangle \\
\left|^{3}A_{2}^{-}\right\rangle 
\end{array}\!\right\} \!=\frac{\left|e_{+}e_{-}\right\rangle -\left|e_{-}e_{+}\right\rangle }{\sqrt{2}}\otimes\left\{ \!\!\begin{array}{c}
\left|\uparrow\uparrow\right\rangle \\
\frac{1}{\sqrt{2}}\left(\left|\uparrow\downarrow\right\rangle +\left|\downarrow\uparrow\right\rangle \right) \text{,} \\
\left|\downarrow\downarrow\right\rangle 
\end{array}\right.
\label{eq:State1}
\end{equation}
where we introduced the $|e_{\pm}\rangle =\frac{1}{\sqrt{2}}\left(|e_{x}\rangle \pm i |e_{y}\rangle \right) $  complex combination of the $e_{\{x,y\}}$ real orbitals. In the ($ee$) electronic configuration, a double degenerate $^1E$ and a non-degenerate $^1A_1$ state appear as
\begin{equation}
\left|^{1}E_\mp\right\rangle =\left.\begin{array}{c}
\left|e_{+}e_{+}\right\rangle \\
\left|e_{-}e_{-}\right\rangle 
\end{array}\right\} \otimes\frac{1}{\sqrt{2}}\left(\left|\uparrow\downarrow\right\rangle -\left|\downarrow\uparrow\right\rangle \right)\label{eq:State2}
\end{equation}
and
\begin{equation}
\left|^{1}A_{1}\right\rangle =\frac{1}{\sqrt{2}}\left(\left|e_{+}e_{-}\right\rangle +\left|e_{-}e_{+}\right\rangle \right)\otimes\frac{1}{\sqrt{2}}\left(\left|\uparrow\downarrow\right\rangle -\left|\downarrow\uparrow\right\rangle \right)\text{.}
\label{eq:State3}
\end{equation}

The optically allowed triplet $^3E$ excited state can be described as an electron promoted from the $a$ to the $e$ orbital in the spin minority channel [see the inclined arrow in Fig.~\ref{fig:NV}(b)] which can be given an $(ae)$ configuration in the hole picture,
\begin{equation}
\left|^{3}E_{\pm}\right\rangle =\left.\!\!\begin{array}{c}
\frac{1}{\sqrt{2}}\left(\left|e_{+}a\right\rangle -\left|ae_{+}\right\rangle \right)\!\\
\frac{1}{\sqrt{2}}\left(\left|e_{-}a\right\rangle -\left|ae_{-}\right\rangle \right)\!
\end{array}\right\} \otimes\left\{ \begin{array}{c}
\left|\uparrow\uparrow\right\rangle \\
\!\!\frac{1}{\sqrt{2}}\left(\left|\uparrow\downarrow\right\rangle +\left|\downarrow\uparrow\right\rangle \right) \text{.}\\
\left|\downarrow\downarrow\right\rangle 
\end{array}\!\right.
\label{eq:State4}
\end{equation}
Beside the triplet state, a double degenerate $^1E^\prime$ state can be constructed as
\begin{equation}
\left|^{1}E_\pm^{\prime}\right\rangle =\left.\begin{array}{c}
\frac{1}{\sqrt{2}}\left(\left|e_{+}a\right\rangle +\left|ae_{+}\right\rangle \right)\\
\frac{1}{\sqrt{2}}\left(\left|e_{-}a\right\rangle +\left|ae_{-}\right\rangle \right)
\end{array}\right\} \otimes\frac{1}{\sqrt{2}}\left(\left|\uparrow\downarrow\right\rangle -\left|\downarrow\uparrow\right\rangle \right) \text{.}\label{eq:State5}
\end{equation}
Both states are Jahn-Teller unstable because a single hole is left in the double degenerate $e$ orbital. We note that $^1E$ in the $(ee)$ electronic configuration is \emph{not} a Jahn-Teller system as closed-shell singlet states are formed in Eq.~\eqref{eq:State2}.

We note that a high energy $A_{1}^{\prime}$ may also exist as follows
\begin{equation}
\left|^{1}A_{1}^{\prime}\right\rangle =\left|aa\right\rangle \otimes\frac{1}{\sqrt{2}}\left(\left|\uparrow\downarrow\right\rangle -\left|\downarrow\uparrow\right\rangle \right) \text{.} \label{eq:State6}
\end{equation}

We particularly focus on the interactions between the singlet states in which an alternative description of the states is useful. The 3 dimensional $|^{1}E\rangle \oplus|^{1}A_{1}\rangle $ can be also expressed by these singlet wavefunctions,
\begin{equation}
\left.\begin{array}{cc}
\left|xx\right\rangle = & \left|e_{x}e_{x}\right\rangle \\
\left|xy\right\rangle = & \frac{1}{\sqrt{2}}\left[\left|e_{x}e_{y}\right\rangle +\left|e_{y}e_{x}\right\rangle \right] \text{,}\\
\left|yy\right\rangle = & \left|e_{y}e_{y}\right\rangle 
\end{array}\right\} \otimes\frac{1}{\sqrt{2}}\left(\left|\uparrow\downarrow\right\rangle -\left|\downarrow\uparrow\right\rangle \right)
\label{eq:PJT1}
\end{equation}
where $|xx\rangle$ is a single Slater-determinant and can be calculated by our HSE06 DFT method. Finally, the $|^{1}E\rangle$ and $|^{1}A_{1}\rangle$  in this basis are
\begin{equation}
\begin{array}{cc}
\left|^{1}E_{x}\right\rangle = & \frac{1}{\sqrt{2}}\left(\left|xx\right\rangle -\left|yy\right\rangle \right)\\
\left|^{1}E_{y}\right\rangle = & \left|xy\right\rangle \\
\left|^{1}A_{1}\right\rangle = & \frac{1}{\sqrt{2}}\left(\left|xx\right\rangle +\left|yy\right\rangle \right)
\end{array}
\label{eq:PJT2}
\end{equation}
that are equivalent to Eqs. \eqref{eq:State2} and \eqref{eq:State3}.

The order of the corresponding levels can be correctly computed by means of configurational interaction or Hubbard Hamiltonian numerical methods~\cite{Delaney2010, Ranjbar2011, Choi2012} that results in $^3A_2$, $^1E$, $^1A_1$, $^3E$, and $^1E^\prime$ levels in ascending order. The $^1A_1^\prime$ level  resides far above that of $^1E^\prime$. This agrees well with the experimental data~\cite{Rogers2008, Kehayias2013} and previous group theory considerations~\cite{Lenef1996, Maze2011, Doherty2011} too. The ZPL energies between the triplets and in-between singlets are at 1.945~eV and 1.19~eV, respectively, are known from PL experiments~\cite{Davies1976, Rogers2008}. The energy gap between the $^3E$ and $^1A_1$ levels [$\Delta$ in Fig.~\ref{fig:NV}(c)] is estimated to be $\sim$0.4~eV from the combination of experimental data and theory on ISC~\cite{Goldman2015a, Goldman2015b}. That would result in $\Sigma\approx0.4$~eV energy gap between the $^1E$ and $^3A_2$ levels. Hubbard Hamiltonian calculations within supercell method~\cite{Choi2012}, that could nearly reproduce the visible and near infared ZPL energies, indeed yielded about 0.4~eV gap between the singlet-triplet levels both in the upper and lower branches.

\subsection{Spin-orbit coupling between the states}

We introduce the spin-orbit coupling (SOC) between the electronic states that is responsible for ISC. The SOC matrix elements between the possible two-particle many-body states can be derived by combining the group theory and the two-component spin-orbit operator on the Slater-determinants of orbitals where 
$\lambda_z$ and $\lambda_\perp$ components of SOC corresponds to the spin-projection conserving and flipping mechanisms, respectively. By following the convention in Ref.~\onlinecite{Doherty2011}, the $^3A_2$ states are linked to $^1A_1$ and $^1E^\prime$ states as follows,
\begin{equation}
\begin{aligned} & \hat{W}=2i\lambda_{z}\left|^{1}A_{1}\right\rangle \left\langle ^{3}A_{2}^{0}\right|+\\
& i\lambda_{\perp}\left[\left|^{1}E_{+}^{\prime}\right\rangle \left\langle ^{3}A_{2}^{+}\right|+\left|^{1}E_{-}^{\prime}\right\rangle \left\langle ^{3}A_{2}^{-}\right|\right]+c.c. \text{,}
\end{aligned}
\label{eq:SOC1}
\end{equation}
where the triplet $m_s=\{0,+,-\}$ manifolds are labeled as a subscript in $^3A_2$.   
The most important conclusion is that $^1E$ in the $(ee)$ electronic configuration is not linked to $^3A_2$. Thus, we seek a possible mechanism that mixes $|^{1}A_{1}\rangle$ and $|^{1}E^\prime\rangle$ characters into $|^{1}E\rangle$, otherwise there would be no any allowed ISC from $|^{1}E\rangle$ to $|^3A_2\rangle$.

\section{Theory on the nature of the shelving singlet state}
\label{sec:4}

In the next sections we derive an approximate wavefunction
of the $|^{1}E\rangle$ including the effects from electron-phonon coupling and many-body electron interaction. First, we derive the pseudo Jahn-Teller effect between the $^1A_1$ and $^1E$ states in Sec.~\ref{sub:4.2}. Next, we determine the dynamic electron-electron correlation between $|^{1}E\rangle$ and $|^{1}E^\prime\rangle$ in Sec.~\ref{sub:4.3} that induces a small but non-negligible dynamic Jahn-Teller effect in $|^{1}E\rangle$. We combine the two effects in Sec.~\ref{sub:4.4}. Despite the small DJT effect, we will demonstrate in Sec.~\ref{sec:5} that only the combination of PJT and DJT accounts for the near infrared PL lineshape of the NV center.

\subsection{Pseudo Jahn-Teller effect between the lowest energy singlet states \label{sub:4.2}}

Since the lowest energy $^1E$ and $^1A_1$ states have different irreducible representations only the symmetry distorting $E$ vibration modes may couple the two states. This effect is known as pseudo Jahn-Teller (PJT) effect in the literature~\cite{bersuker2012vibronic, bersuker2013jahn}. We work out the PJT Hamiltonian in the basis of  $\left|xx\right\rangle$, $\left|xy\right\rangle$, and $\left|yy\right\rangle$ wavefunctions (see Eqs.~\eqref{eq:PJT1} and \eqref{eq:PJT2}). By assuming an electronic gap of $\Lambda_e$ between the $^1E$ and $^1A_1$ before turning on the electron-phonon interaction and setting the energy of $^1E$ to zero we arrive at
\begin{equation}
\begin{aligned} & \hat{H}= \underset{{\textstyle =\hat{H}_{e}}}{\underbrace{\frac{\Lambda_e}{2}\begin{pmatrix}1 & 0 & 1\\
		0 & 0 & 0\\
		1 & 0 & 1
		\end{pmatrix}}} + \\
& \underset{{\textstyle =\hat{H}_{\mathrm{osc.}}}}{\underbrace{\hbar\omega_{E}\left(\sum_{\alpha\in \{x,y\}}a_{\alpha}^{\dagger}a_{\alpha}+1\right)}}+\underset{{\textstyle =\hat{H}_{\mathrm{PJT}}}}{\underbrace{\widetilde{F}\left(\hat{\sigma}_{z}\hat{x}-\hat{\sigma}_{x}\hat{y}\right) \text{,}}}
\end{aligned}
\label{eq:PJT4}
\end{equation}
where $\hat{H}_{e}$, $\hat{H}_{\mathrm{osc.}}$ and $\hat{H}_{\mathrm{PJT}}$ are the electronic, harmonic oscillator and linear PJT Hamiltonian, respectively. We note that $\Lambda_e$ is not exactly the ZPL energy ($\Lambda$) between the singlets because that will be corrected by the vibronic energies coming from the electron-phonon interactions. The $\hat{\sigma}_{z}$ and $\hat{\sigma}_{x}$ operators
\begin{equation}
\hat{\sigma}_{z}=\begin{pmatrix}1 & 0 & 0\\
0 & 0 & 0\\
0 & 0 & -1
\end{pmatrix}\qquad\hat{\sigma}_{x}=\frac{1}{\sqrt{2}}\begin{pmatrix}0 & 1 & 0\\
1 & 0 & 1\\
0 & 1 & 0
\end{pmatrix}
\label{eq:PJT5}
\end{equation}
are matrix representation of the $L=1$ angular momentum pointing
to the electronic degree of freedom in the PJT interaction. The $E$ vibration mode is described by $a_{x,y}^{\dagger}$ creation and $a_{x,y}$ annihilation operators with $\omega_{E}$ frequency. For the sake of simplicity we use dimensionless coordinates where $\hat{x}=(a_{x}^{\dagger}+a_{x})/\sqrt{2}$
and $\hat{y}=(a_{y}^{\dagger}+a_{y})/\sqrt{2}$. We show below that other states contribute to the electron-phonon interaction in the $^1E$ state.

\subsection{Dynamic electron-electron correlation between the $|^{1}E\rangle $ and $|^{1}E^\prime\rangle$ states and the appearance of dynamic Jahn-Teller effect\label{sub:4.3}}

The electron-electron correlation is possible among the many-body states with the same irreducible representation as described in Eqs. \eqref{eq:State1}-\eqref{eq:State5}.Thus it might be possible that the $|^{1}A_{1}\rangle$
and $|^{1}A_{1}^{\prime}\rangle$  correlate at a certain
degree as well as the $|^{1}E\rangle$ and $|^{1}E^{\prime}\rangle$
do that similarly. We focus on the mixture of  $|^{1}E\rangle$ and $|^{1}E^{\prime}\rangle$ as this would allow $\Gamma_\perp = \Gamma_\pm + \Gamma_\mp$ ISC process between $|^{1}E\rangle$ and $|^{3}A_2^\pm\rangle$. The mixing coefficient $C$ describes a multi-determinant singlet state ($\left|^{1}\bar{E}\right\rangle$) as
\begin{equation}
\left|^{1}\bar{E}\right\rangle =C\left|^{1}E\right\rangle + \sqrt{1-C^2}\left|^{1}E^{\prime}\right\rangle \text{,} 
\label{eq:Corr1}
\end{equation}
where $C$ can be chosen to be real without losing the generality. Since $\left|^{1}E^{\prime}\right\rangle$ is an $E\otimes e$ DJT system $\left|^{1}\bar{E}\right\rangle$ carries a DJT character by the extent of $(1-C^2)$. The DJT Hamiltonian of $\left|^{1}E^{\prime}\right\rangle$ is
\begin{equation}
\hat{H}_{\mathrm{JT}}= F\big(\bar{\sigma}_{z}\hat{X}-\bar{\sigma}_{x}\hat{Y}\big) \text{,}
\label{eq:Corr2}
\end{equation}
where the electronic degree of freedom is expressed by $\bar{\sigma}_{z}$
and $\bar{\sigma}_{x}$ Pauli matrices spanning the two dimensional $|^{1}E^{\prime}\rangle $ space that can be written as
$\bar{\sigma}_{z}=|^{1}E_{x}^{\prime}\rangle \langle ^{1}E_{x}^{\prime}|-|^{1}E_{y}^{\prime}\rangle \langle ^{1}E_{y}^{\prime}|$
and $\bar{\sigma}_{x}=|^{1}E_{x}^{\prime}\rangle \langle ^{1}E_{y}^{\prime}|+|^{1}E_{y}^{\prime}\rangle \langle ^{1}E_{x}^{\prime}|$. The effective DJT Hamiltonian in the basis of Eq.~\eqref{eq:PJT1} used for the PJT Hamiltonian is
\begin{equation}
\hat{H}_{\mathrm{JT}}^{\mathrm{eff}}=(1-C^2)F\big(\bar{\sigma}_{z}\hat{X}-\bar{\sigma}_{x}\hat{Y}\big)
\label{eq:Corr4}
\end{equation}
with 
\begin{equation}
\bar{\sigma}_{z}=\frac{1}{2}\begin{pmatrix}1 & 0 & -1\\
	0 & 0 & 0\\
	-1 & 0 & 1
\end{pmatrix}\quad\bar{\sigma}_{x}=\frac{1}{\sqrt{2}}\begin{pmatrix}0 & 1 & 0\\
1 & 0 & -1\\
0 & -1 & 0
\end{pmatrix} \text{.}
\label{eq:Corr5}
\end{equation}

\subsection{The effective electron-phonon Hamiltonian for the shelving singlet state and the vibronic wavefunctions\label{sub:4.4}}

The PJT effect was developed for $|{}^{1}E\rangle$, however, we learned in Sec.~\ref{sub:4.3} that the lowest energy singlet is rather $\left|^{1}\bar{E}\right\rangle$ that will reduce $\hat{H}_{\mathrm{PJT}}$ by $C^2$. Furthermore, one can estimate that the electron-phonon coupling in PJT ($\tilde{F}$) and in DJT ($F$) has a relation of $\tilde{F}\approx 2F$. This relation can be envisioned from the two-particle $|xx\rangle$ singlet wavefunction associated with its electron-phonon coupling $\tilde{F}$ in which both orbitals are Jahn-Teller unstable whereas only a single orbital is Jahn-Teller unstable in $(ae)$ electronic configuration associated with its electron-phonon coupling $F$. As a consequence, the Jahn-Teller effect is twice stronger in $|xx\rangle$ than that in $(ae)$ electronic configuration. The final effective electron-phonon Hamiltonian of the shelving singlet state is
\begin{equation}
\hat{H}^{\mathrm{eff}}_{\mathrm{el-ph}}=C^2 2 F \left(\hat{\sigma}_{z}\hat{X}-\hat{\sigma}_{x}\hat{Y}\right) + (1-C^2)F\big(\bar{\sigma}_{z}\hat{X}-\bar{\sigma}_{x}\hat{Y}\big) \text{.}
\label{eq:Corr6}
\end{equation}

The full Hamiltonian for the $^1\widetilde{E}\oplus ^1\widetilde{A}_1$ system is
\begin{equation}
\hat{H}=\hat{H}_{\mathrm{e}}+\hat{H}_{\mathrm{osc.}}+\hat{H}^{\mathrm{eff}}_{\mathrm{el-ph}}\text{,}
\label{eq:nPh2}
\end{equation}
which results in the following $\widetilde{\Psi}$ vibronic wavefunctions in the expansion of $E$ phonon modes as follows,
\begin{equation}
\begin{aligned}\bigl|\widetilde{\Psi}\bigr\rangle  & =\sum_{n,m}^{\infty}\left(c_{nm}^{xx}\left|xx\right\rangle \otimes\left|nm\right\rangle +c_{nm}^{xy}\left|xy\right\rangle \otimes\left|nm\right\rangle +\right.\\
& \left.c_{nm}^{yy}\left|yy\right\rangle \otimes\left|nm\right\rangle \right)
\end{aligned}
\label{eq:nPh1}
\end{equation}
where we limit the expansion in the Born-Oppenheimer basis ($\left|nm\right\rangle =\frac{1}{\sqrt{nm}}(a_{x}^{\dagger})^{n}(a_{y}^{\dagger})^{m}\left|00\right\rangle$) up to 10 phonon limit $n+m\leq10$ which is numerically convergent in our particular case. We span the electronic degrees of freedom with $\left|xx\right\rangle$, $\left|xy\right\rangle$, $\left|yy\right\rangle$ as we defined in 
Eq.~\eqref{eq:PJT2}. In this form one can express the combined $|{}^{1}\widetilde{A}_{1}\rangle \oplus |{}^{1}\widetilde{E}_{\pm}\rangle$ states which may transform as $E$, $A_1$ and $A_2$. The $\widetilde{A}_2$ vibronic states do not play a significant role, thus we only show the expressions for the $^1\widetilde{E}_\pm$ and $^1\widetilde{A}_1$ vibronic states as follows
\begin{widetext}
\begin{subequations}
	\begin{align}&
\left|^{1}\widetilde{E}_{\pm}\right\rangle =\sum_{i=1}^{\infty}\left[c_{i}\left|^{1}E_{\pm}\right\rangle \otimes\left|\chi_{i}\left(A_{1}\right)\right\rangle +d_{i}\left|^{1}A_{1}\right\rangle \otimes\left|\chi_{i}\left(E_{\pm}\right)\right\rangle +f_{i}\left|^{1}E_{\mp}\right\rangle \otimes\left|\chi_{i}\left(E_{\mp}\right)\right\rangle +g_{i}\left|^{1}E_{\pm}\right\rangle \otimes\left|\chi_{i}\left(A_{2}\right)\right\rangle \right]
	\label{eq:Epolaron}
	\\&
	\left|^{1}\widetilde{A}_{1}\right\rangle =\sum_{i=1}^{\infty}\left[c_{i}^{\prime}\left|^{1}A_{1}\right\rangle \otimes\left|\chi_{i}\left(A_{1}\right)\right\rangle +\frac{d_{i}^{\prime}}{\sqrt{2}}\left(\left|^{1}E_{+}\right\rangle \otimes\left|\chi_{i}\left(E_{-}\right)\right\rangle +\left|^{1}E_{-}\right\rangle \otimes\left|\chi_{i}\left(E_{+}\right)\right\rangle \right)\right]
	\label{eq:Epolaron2}
	\end{align}
	\label{eq:spin-orbit}
\end{subequations} 
\end{widetext}
that govern the shape of the phonon
sideband in the optical spectra. We label the symmetry adapted vibrational wavefunctions, e.g., 
$\left|\chi_{1}\left(A_{1}\right)\right\rangle =\left|00\right\rangle$, or in general, by 
$\left|\chi_{i}\left(\dots\right)\right\rangle$ in the rest of the paper.

We note that the $g_i$ coefficients are generally tiny and will be ignored. On the other hand, the non-zero $d_i$ and $c_i^\prime$ ($f_i$ and $d_i^\prime$) coefficients drive the $\Gamma_z$ ($\Gamma_\perp$) ISC process, and they are also responsible for the shape of PL spectrum of the singlets.

We further note that the analysis of strain dependence of the singlet states requires to extend the effective electron-phonon Hamiltonian (Eq.~\eqref{eq:Corr6}) with a strain Hamiltonian that have similar matrix elements like the effective electron-phonon Hamiltonian Hamiltonian that may explain the significant strain interaction of the $^1\widetilde{E}_\pm$ level~\cite{Rogers2015}. A detailed investigation on the interaction of the $^1\widetilde{E}_\pm \oplus ^1\widetilde{A}_1$ vibronic levels with the strain is out of the scope of this study, and we rather concentrate on understanding the role of the singlet vibronic states in the intersystem crossing processes of NV center.

\section{Application of \emph{ab initio} results on the theoretical models\label{sec:5}}

We first estimate the parameters in the developed electron-phonon Hamiltonian from \emph{ab initio} DFT calculations for the singlet shelving state in Sec.~\ref{sub:5.1} and present vibronic electronic structure. We then apply the resulting vibronic wavefunctions to calculate the PL and absorption spectra for the singlets that will verify our methodology in Sec~\ref{sub:5.3}. Finally, we determine the ISC rates between $|^{1}\widetilde{E}\rangle$ into $|^{3}A_2\rangle$ in Sec.~\ref{sub:5.4} that show good agreement with the experimental data, including the temperature dependence.

\subsection{Derivation of the parameters of the electron-phonon Hamiltionian of the singlet states from DFT calculations and the resulting vibronic levels\label{sub:5.1}}

The true many-body singlet eigenstates of NV center cannot be exactly described by Kohn-Sham DFT methods. Nevertheless, the closed-shell $|xx\rangle$ state can be expressed. Furthermore, the open-shell  $|a_{\uparrow}a_{\downarrow}e_{y\uparrow}e_{x\downarrow}\rangle$ can be also calculated by $\Delta$SCF Kohn-Sham DFT method. We calculated these singlet states by constraining the symmetry into $C_{3v}$ started from the optimized $^3A_2$ ground state geometry. We found that the geometry did not practically change in the geometry optimization procedure of these singlet states which strongly hints that the $A_1$ phonons do not contribute to the change of the geometry of the true $^1A_1$ and $^1E$ states w.r.t.\ that of $^3A_2$ ground state, thus the symmetry-breaking $E$ phonons play a role.

\begin{figure*}
	\includegraphics[width=\textwidth]{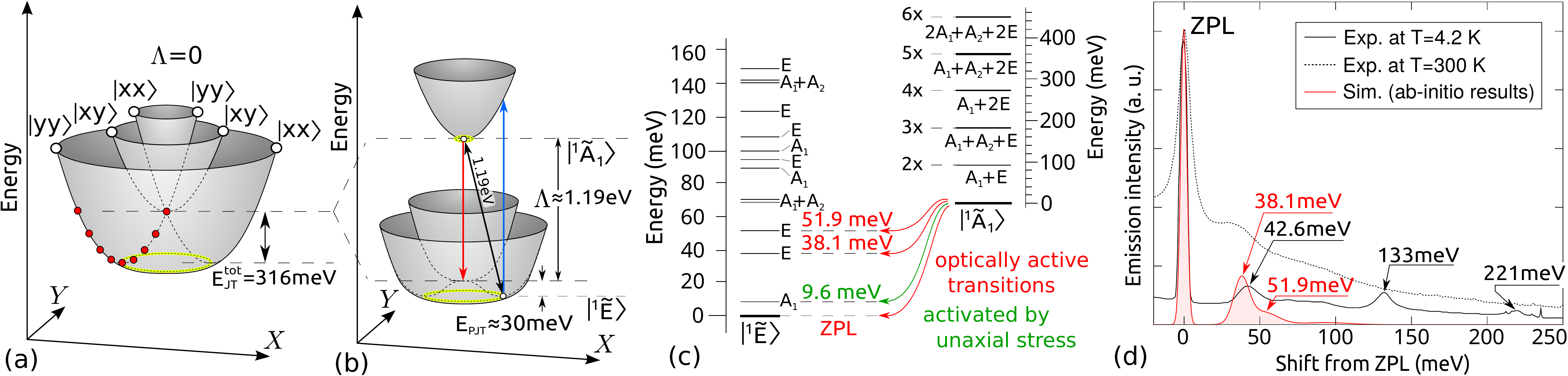}
	\caption{\label{fig:PJT}(a) Jahn-Teller nature of the $|^{1}A_{1}\rangle\oplus|^{1}E\rangle$ states as can be calculated by means of Kohn-Sham DFT which corresponds to $\Lambda_e$=0~eV. The resulted Jahn-Teller energy 
$E_\text{JT}$=316~meV. (b) After $\Lambda_e\approx$1.13~eV is switched on the resulting Jahn-Teller energy is $E_\text{PJT}\approx$30~meV according to the solution of Eq.~\eqref{eq:nPh2} that results in $\Lambda$=1.19~eV ZPL energy. (c) The vibronic levels of $|^{1}\widetilde{E}\rangle$ and $|^{1}\widetilde{A}_{1}\rangle$. The selection rules for the photoluminescence spectrum is indicated. Here the ZPL energy of 1.19~eV between $|^{1}\widetilde{E}\rangle$ and $|^{1}\widetilde{A}_{1}\rangle$ is not scaled for the sake of clarity. (d) Experimental photoluminescence spectrum of the singlets at low (black solid line) and room (dotted black line) temperatures compared to the simulated spectrum from \emph{ab initio} solution (red curve). We note that the experimental spectra show a substantial and minor background at room and cryogenic temperature, respectively. The simulation curve does not include background signal. The ZPL energy is now set to zero in order to easily read out the position of vibration features in the spectrum. We used 2~meV, 5~meV, and 10~meV gaussian smearing for the linewidth of the ZPL, first and second vibronic emissions, respectively, where the width of the ZPL and vibration bands are read out from the experimental spectrum recorded at cryogenic temperature.}	
\end{figure*}

In particular, the $|xx\rangle$ is useful to derive the electron-phonon Hamiltonian of the singlet states as this state appears in Eq.~\eqref{eq:Corr6}. 
By allowing free movement of atoms in the geometry optimization procedure, the $|xx\rangle$ state spontaneously reconstructs to $C_{1h}$ symmetry with a $E_\text{JT}$ Jahn-Teller energy of 316~meV and an effective phonon mode $\hbar \omega_E$=66~meV where the latter is the solution of the quasi-Harmonic oscillator in the APES of $|xx\rangle$. This $|xx\rangle$ state does not contain the strong electron correlation, e.g., the electronic gap $\Lambda_e\approx\Lambda$=1.19~eV between the $^1E$ and $^1A_1$ states. It can be easily shown by perturbation theory (see Appendix \ref{app:A}) that $E_\text{JT}$ will be damped by $\Lambda_e$ in $^1E$ state [c.f., Figs.~\ref{fig:PJT}(a) and (b)]. The exact results come from solving the full electron-phonon Hamiltonian in Eq.~\eqref{eq:Corr6}. That requires to identify parameter $C^2$ which is associated with the contribution of $^1E^\prime$ state in $^1\bar{E}$ state.   

We estimate parameter $C^2$ from the character of the Kohn-Sham wavefunctions of $|xx\rangle$ in the $C_{1h}$ APES global minimum. Although we calculated $|xx\rangle$ by non-spinpolarized DFT, the symmetry distortion may result in the contribution of the $a$ Kohn-Sham orbital in the two-particle wavefunctions. By labeling the Kohn-Sham orbital in the distorted geometry by $\xi$, and the contribution of $e_x$ and $a$ orbitals by $p$ and $s$, respectively, one arrives at 
\begin{equation}
\begin{aligned}\left|\xi\xi\right\rangle =\left[p\left|e_{x}\right\rangle +s\left|a\right\rangle \right]\left[p\left|e_{x}\right\rangle +s\left|a\right\rangle \right]=\\
p^{2}\left|e_{x}e_{x}\right\rangle +\sqrt{2}ps\underset{{\textstyle \left|^{1}E_{x}^{\prime}\right\rangle }}{\underbrace{\frac{\left|ae_{x}\right\rangle +\left|ea\right\rangle }{\sqrt{2}}}}+s^{2}\underset{{\textstyle \left|^{1}A_{1}^{\prime}\right\rangle }}{\underbrace{\left|aa\right\rangle}}\text{,}
\end{aligned}
\label{eq:methods2}
\end{equation}
where $(1-C^2)=2p^2s^2$ can be read out. By using projector operators such as $s=\langle \xi|a\rangle$ we find that the contribution of $^1A_1^\prime$ is minor and can be neglected, however, $(1-C^2)$=0.1 is significant, and explains the $\Gamma_\perp$ ISC processes. 

This correlation effect also brings a DJT effect to the electron-phonon Hamiltonian. Thus, the calculated APES of $|xx\rangle$ contains both PJT and DJT effects that should be separated. This is established in Eq.~\eqref{eq:Corr6} where the corresponding Jahn-Teller energy is
\begin{equation}
E_{\text{JT}}=\frac{\left(C^{2} 2F+ (1-C^{2})F \right)^{2}}{2\hbar\omega_E}\text{.}
\label{eq:methods3}
\end{equation}
By using the previously determined $E_{\text{JT}}$, $C^2$ and $\hbar \omega_E$ from DFT APES calculations of $|xx\rangle$, we obtain $F$=102.47~meV. Thus, we have all the parameters but $\Lambda_e$ to build up the full electron-phonon Hamiltonian of the singlet states.

The measured ZPL energy between the singlet states is 1.19~eV which is the energy difference between the vibronic ground states of the singlets. We fit the value of $\Lambda_e$ to obtain the experimental ZPL energy after diagonalizing the full electron-phonon Hamiltonian which resulted in $\Lambda_e$=1129.4~meV. The vibronic levels are depicted in Fig.~\ref{fig:PJT}(c) where the corresponding coefficients of the vibronic states are listed in Table~\ref{tab:coeffs} in Appendix~\ref{app:vibwfs}. One can see interesting features in the calculated vibronic spectra: (i) the vibronic levels of $|^{1}\widetilde{E}\rangle$ is very far from the solution of a quasi-harmonic oscillator and shows up a very complex feature; (ii) the vibronic levels of $|^{1}\widetilde{A}_{1}\rangle$ are equidistant, however, the PJT effect will increase the effective phonon mode of 66~meV to 91.8~meV. 

It is important to highlight the complex interplay between the PJT and DJT effects in the final vibronic spectrum of $|^{1}\widetilde{E}\rangle$. Although, DJT effect is damped by $(1-C^2)^2$ factor, still it changes the spectrum at $\approx$45~meV, and results in two split $E$ levels that would not be otherwise there. In addition, it changes the character of these vibronic wavefunctions, so that it increases the optical transition dipole moments with the ground state vibronic state of $|^{1}\widetilde{A}_{1}\rangle$. Our results demonstrate that electron-electron correlation effect combined with electron-phonon couplings of different nature and involving three electronic states can only fully describe the electron-phonon system of the singlet states in NV center. We will show that this complex nature can only account for the measured optical spectra and ISC rates.

\subsection{Vibronic sideband of the 1.19-eV photoluminescence and absorption spectrum}
\label{sub:5.3}

We discuss now the PL and absorption spectrum of the singlets. Several features of these spectrum can be understood by our vibronic wavefunctions that verify our method in the calculation of the ISC rates. The relative optical transition dipole strength between the vibronic $^{1}\widetilde{E}$ and $^{1}\widetilde{A}_{1}$ ground states is derived in Appendix \ref{app:B}.

We first discuss the luminescence spectrum at low temperatures which is a radiative decay between the $^{1}\widetilde{A}_{1}$ ground state and the vibronic ground and $n$th excited states of $^{1}\widetilde{E}_{x}$ [see Eq.~\eqref{eq:Lum7}], i.e.,
\begin{equation}
I\left(^{1}\widetilde{A}_{1}\rightarrow{}^{1}\widetilde{E}^{(n)}\right)=\left|\left\langle ^{1}\widetilde{A}_{1}\right|\hat{d}_{x}\left|^{1}\widetilde{E}^{(n)}\right\rangle \right|^{2} \text{.}
\label{eq:Opt}
\end{equation}
We found from direct calculation of the intensities in Eq.~\eqref{eq:Opt} that the optical transition to the first vibronic $A_1$ state of $^{1}\widetilde{E}$ state is not allowed. However, there is a significant optical transition dipole toward the split $E$ vibronic states around 45~meV. After switching off the small DJT effect in the electron-phonon Hamiltonian only a single $E$ mode appears with a smaller optical transition dipole moment. This clearly demonstrates that the small DJT effect does play an important role in understanding the optical features of the singlet states. The simulated PL spectrum from \emph{ab initio} wavefunctions is shown in Fig.~\ref{fig:PJT}(d) that can be directly compared to the low temperature experimental PL spectrum~\cite{Rogers2008}. Clearly, the broad feature with the maximum intensity at $\approx$43~meV can be reproduced (red curve). We find that the broad feature consists of two close-level vibronic excited states  [see red text in Fig.~\ref{fig:PJT}(c)]. The experimental intensity and the shape of this broad feature can be well reproduced by invoking our electron-phonon Hamiltonian (red curve). Our theory does not account for the features at 133~meV and 221~meV. These features seem to disappear at room temperature PL spectrum, thus we conclude that they may not belong to NV center. Our theory is further supported by an uniaxial stress experiment on the PL spectrum which showed up the existence of a forbidden state at $\approx$14~meV~\cite{manson2010optically}. This can be naturally explained by our calculated $A_1$ vibronic excited state [see green text in Fig.~\ref{fig:PJT}(c)]. This $A_1$ state will play an important role in the temperature dependence of the ISC rate where $\approx$16~meV phonon mode was deduced from the temperature dependent ISC rate measurements in non-stressed diamond samples~\cite{Robledo2011} that should be identical with the optically forbidden vibronic mode.   

\begin{figure}[ht] 
	\includegraphics[width=240pt]{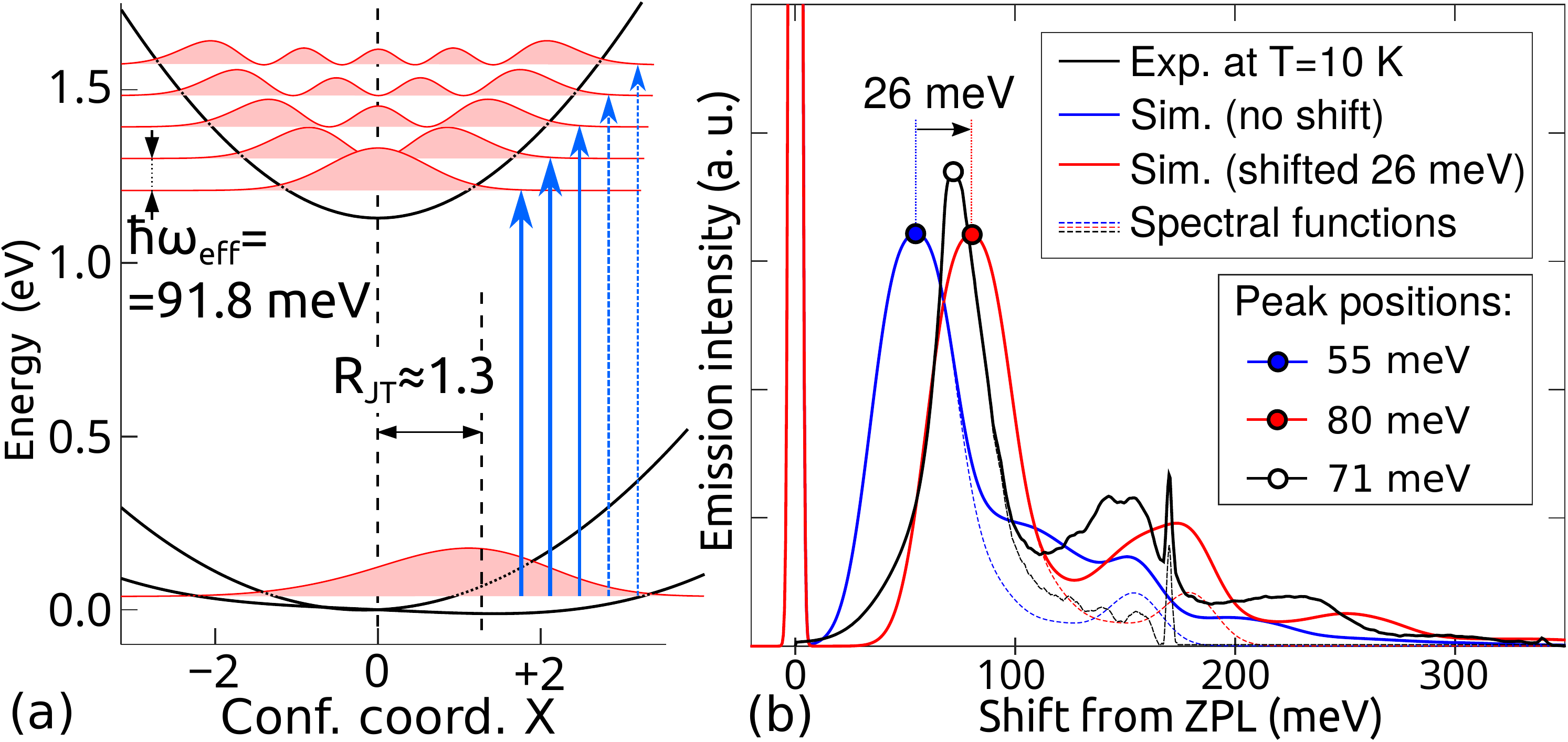} 
	\caption{\label{fig:HR}(a) Calculated APES of singlet states of NV center including the pseudo and dynamic Jahn-Teller effects simultaneously but with $Y=0$ constraint. Zero value at the configuration coordinates corresponds to $C_{3v}$ symmetry. The localization of the wavefunctions is depicted. We note that the potential energy is not axial symmetric as explained in the text. (b) 
Low temperature experimental absorption spectrum (black solid curve) and the deduced spectral function (black dotted curve) from Ref.~\onlinecite{Kehayias2013} are compared with the calculated absorption spectra (solid blue and red curves) and spectral functions (dotted blue and red curves) with using the Huang-Rhys theory based on the APES of singlet states. We applied 1.5~meV gaussian smearing on the theoretical lineshapes.}
\end{figure}

Now we turn to the absorption spectrum which is very different from the PL spectrum [c.f. Fig.~\ref{fig:PJT}(d) and Fig.~\ref{fig:HR}(b)]. We can explain this feature by the presence of simultaneous PJT and DJT effects. The PJT and the DJT effects are separately create an axial symmetric APES about the symmetry axis of the defect, however, the DJT will create a barrier energy for the free rotation about the symmetry axis in the PJT APES. This can be readily observed by comparing Eqs.~\eqref{eq:PJT5} and \eqref{eq:Corr5} corresponding to PJT and DJT effects, respectively, which differently combine the wavefunctions upon the same distortion.  In the absorption process we assume that the photon absorption is a faster process than the quantum mechanical tunneling between the global minima of APES, i.e., the $Y$-axis is frozen in Fig.~\ref{fig:PJT}(b) which leads to the APES in Fig.~\ref{fig:HR}(a). The ground state vibronic wavefunction becomes localized in one of the APES valleys at a distance of about 1.3 from the $C_{3v}$ symmetry position. For the $^1\widetilde{A}_1$ excited state the DFT APES predicted $\hbar\omega_E$=66.1~meV, however, the solution of the full electron-phonon Hamiltonian revealed us that PJT effect increases this energy to 91.8~meV. Therefore, we also created the corresponding harmonic APES [see Fig.~\ref{fig:HR}(a)] and employed the Huang-Rhys theory to produce, where $R$=1.3 results in an $S\approx$0.84 Huang-Rhys factor. Finally, the absorption spectrum is produced [Fig.~\ref{fig:HR}(a)] with the $\hbar\omega_E$=66.1~meV (blue curve) and also shifted one, $\hbar\omega_E$=91.8~meV (red curve). We emphasize that these effective phonon frequencies create high peaks at 55~meV and 81~meV in the absorption spectrum with respect to the position of the ZPL. The latter is much closer to the experimental~\cite{Kehayias2013} peak at 71~meV [see Fig.~\ref{fig:HR}(b)].  It can be concluded that the electron-phonon Hamiltonian based effective phonon frequency is slightly overestimated but the calculated broad features in the absorption phonon sideband are in agreement with the observed ones. The corresponding spectral functions [dotted curves in Fig.~\ref{fig:HR}(b)] either deduced from experimental spectrum (black dotted line) or calculated (blue and red dotted lines) are also shown. We note that the very sharp feature at high energy in the experimental spectrum is associated with the quasi-local nitrogen-carbon vibration modes~\cite{Kehayias2013}. Our theory does not account to that feature that would require the exact calculation of APES of the singlet states. On the other hand, the PJT-DJT theory (red curve) brings the results close to the experimental values and explains the upward shift in the first characteristic phonon peak w.r.t.\ that of the triplets ($\approx$64~meV).  

We conclude from these results that the combination of PJT and DJT effects accounts for the observed asymmetry in the lineshape of the emission and absorption spectrum. In the PL process, the selection rules are dictated by the dynamic motion of ions combined with the electron wavefunction which results in an optically forbidden transition that becomes visible under uniaxial stress. On the other hand, this dark vibronic state of the shelving singlet can play an important role in the ISC process. In the absorption process, the dynamics of ions is frozen thus it shows up in the optical spectrum, and optical transition to all of the vibronic states of the upper singlet state is allowed. The dynamics of ions can be slower than the absorption of the photons because of the small but non-negligible DJT effect which produces an energy barrier for the motion of ions. This leads to the large asymmetry in the corresponding phonon sidebands. Furthermore, PJT explains the enhanced effective phonon energy in the absorption spectrum of the singlets with respect to that in the optical spectrum of the triplets. These results verify our theory on the singlets and serve as a good base to study the ISC process between the shelving singlet state and the triplet ground state.

\subsection{Theory and \emph{ab initio} results on the ISC process toward the ground state}
\label{sub:5.4}

We determine the ISC rates from $|^{1}\widetilde{E}\rangle$ to $|^{3}A_2\rangle$ based on the vibronic states calculated from DFT wavefunctions and potentials. The ISC process is a spin-orbit driven scattering of the electron that is mediated by phonons for energy conversation, as the expected energy difference between $^{1}\widetilde{E}$ to $^{3}A_2$ levels is several orders of magnitude larger than the spin-orbit energy. In other words, the electron is scattered to the vibration levels ($\left\langle ...\right|$) of the $^3A_2$ ground state. As we discussed previously, the emission or absorption of $A_1$ phonons is minor in the process, thus we rely on the contribution of the $E$ phonons that are responsible for the PJT and DJT effects. The ISC rate can be calculated using the Fermi golden-rule and assuming the strength of spin-orbit coupling does not change significantly upon the motion of ions in the process. This theory was developed for the ISC process between $^{3}E$ and $^{1}A_1$ states in the upper branch by Goldman and co-workers~\cite{Goldman2015a, Goldman2015b} that we developed further to take into account the vibronic nature of $^{3}E$ state, i.e., $^{3}\widetilde{E}$ caused by DJT effect~\cite{Thiering2017}. By applying this theory to the $\Gamma_z$ ISC rate between $|{}^{1}\widetilde{E}\rangle$ and vibration state of $^3A_2$ we arrive at
\begin{equation}
\begin{aligned} & \Gamma_{z}=\frac{2\pi C^2}{\hbar}\sum_{\left|...\right\rangle }\left|\left\langle ...\right|\otimes\left\langle ^{3}A_{2}^{0}\right|\hat{W}\left|^{1}\widetilde{E}\right\rangle \right|^{2}\delta\left(\Sigma-E\left(\left|...\right\rangle \right)\right)\\
& =\frac{2\pi C^2}{\hbar}\sum_{i}^{\infty}4\lambda_{z}^{2} d_{i}^{2}\underset{{\textstyle \approx S_{E}^{(n_{i})}\left(\Sigma\right)}}{\underbrace{\left|\langle...\left|\chi_{i}\left(E_\pm\right)\right\rangle \right|^{2}\delta\bigl(\Sigma-n_{i}\hbar\omega_{E}\bigr)}}\\
& \approx\frac{8\pi\lambda_{z}^{2}C^2}{\hbar}\sum_{i}^{\infty} d_{i}^{2}S_{E}^{(n_{i})}\left(\Sigma\right)=\frac{8\pi\lambda_{z}^{2}C^2}{\hbar}F_{E}\left(\Sigma\right)
\end{aligned}
\label{eq:gamma1}
\end{equation}
where the summation over all vibration wavefunctions of $^3A_2$ collapses to  the number of $|\chi_{i}(E_\pm)\rangle$ vibration modes in the phonon overlap integral. Here, $d_i$ coefficient is responsible to the contribution of electronic $^1A_1$ state in $|{}^{1}\widetilde{E}\rangle$ that is connected to $^3A_2^0$ by $\lambda_z$. 
Now the energy conservation law is $\Sigma=n_i\times\hbar\omega_E$ for some $n_i$ ($n_i$ is the phonon index of the $i$th $\left|\chi_{i}\left(E_{\pm}\right)\right\rangle$ vibronic function). Here $S_{E}$ is the phonon overlap spectral function and $F_E$ is the modulated phonon overlap function caused by PJT effect. $\Sigma$ is the ZPL energy between  $|{}^{1}\widetilde{E}\rangle$ and $^3A_2$ [see Fig.~\ref{fig:NV}(c)]. So far we used effective phonon energies with discrete quantum levels but this would lead to often a zero overlap in $F_E$. In reality, the diamond phonons interact with the quasi-local vibration modes found in PJT and DJT effects that can be described as a smearing of the energy spectrum of the quasi-local vibration modes. In order to incorporate this effect, we autoconvolute the electron-phonon modes $n_{i}$ times by defining the following recursive formula,
\begin{equation}
S_{E}^{(n)}\left(x\right)=\left(S_{E}^{(n-1)}\ast S_{E}\right)\left(x\right)\qquad S_{E}^{(0)}\left(x\right)=\delta\left(x\right) \text{,}
\label{eq:conv}
\end{equation}
where "$\ast$" labels the convolution, and $\delta(x)$ is the Dirac delta function. Similar considerations have been applied recently (see the Supplemental Material in Ref.~\onlinecite{Kehayias2013}).

Beside $\Gamma_z$ ISC processes, the $\Gamma_\pm$ and $\Gamma_\mp$ ISC processes can take place governed by $\lambda_\perp$ because of the contribution of $^1E^\prime$ in 
$|{}^{1}\widetilde{E}\rangle$. By applying the Fermi golden-rule again, we arrive at
\begin{equation}
\begin{aligned} & \Gamma_{\pm}=\frac{2\pi (1-C^2)}{\hbar}\sum_{\left|...\right\rangle }\left|\left\langle ...\right|\otimes\bigl\langle^{3}A_{2}^{\pm}\bigr|\hat{W}\bigl|{}^{1}\widetilde{E}\bigr\rangle\right|^{2}\delta\left(\Sigma-E\left(\left|...\right\rangle \right)\right)\\
& =\frac{2\pi (1-C^{2})}{\hbar}\sum_{i}^{\infty}\lambda_{\perp}^{2}c_{i}^{2}\left|\langle...\left|\chi_{i}\left(A_{1}\right)\right\rangle \right|^{2}\delta\left(\Sigma-n_{i}\hbar\omega_{E}\right)\\
& \approx\frac{2\pi (1-C^{2})\lambda_{\perp}^{2}}{\hbar}\sum_{i}^{\infty}c_{i}^{2}S_{E}^{(n_{i})}\left(\Sigma\right)=\frac{2\pi (1-C^{2})\lambda_{\pm}^{2}}{\hbar}F_{E}^{\prime}\left(\Sigma\right)
\end{aligned}
\label{eq:gamma2}
\end{equation}
and
\begin{equation}
\begin{aligned} & \Gamma_{\mp}= \frac{2\pi (1-C^2)}{\hbar} \sum_{\left|...\right\rangle }\left|\left\langle ...\right|\otimes\bigl\langle^{3}A_{2}^{\pm}\bigr|\hat{W}\bigl|{}^{1}\widetilde{E}\bigr\rangle\right|^{2}\delta\left(\Sigma-E\left(\left|...\right\rangle \right)\right)\\
& =\frac{2\pi (1-C^{2})}{\hbar}\sum_{i}^{\infty}\lambda_{\perp}^{2}f_{i}^{2}\left|\langle...\left|\chi_{i}\left(E_{\mp}\right)\right\rangle \right|^{2}\delta\left(\Sigma-n_{i}\hbar\omega_{E}\right)\\
& \approx\frac{2\pi (1-C^{2})\lambda_{\perp}^{2}}{\hbar}\sum_{i}^{\infty}f_{i}^{2}S_{E}^{(n_{i})}\left(\Sigma\right)=\frac{2\pi (1-C^{2})\lambda_{\perp}^{2}}{\hbar}F_{E}^{\prime\prime}\left(\Sigma\right) \text{,}
\end{aligned}
\label{eq:gamma3}
\end{equation}
where $F_E^\prime$ and $F_E^{\prime\prime}$ are the corresponding phonon overlap spectral functions caused by DJT effect. 

\begin{figure}[ht] 
	\includegraphics[width=7.5cm]{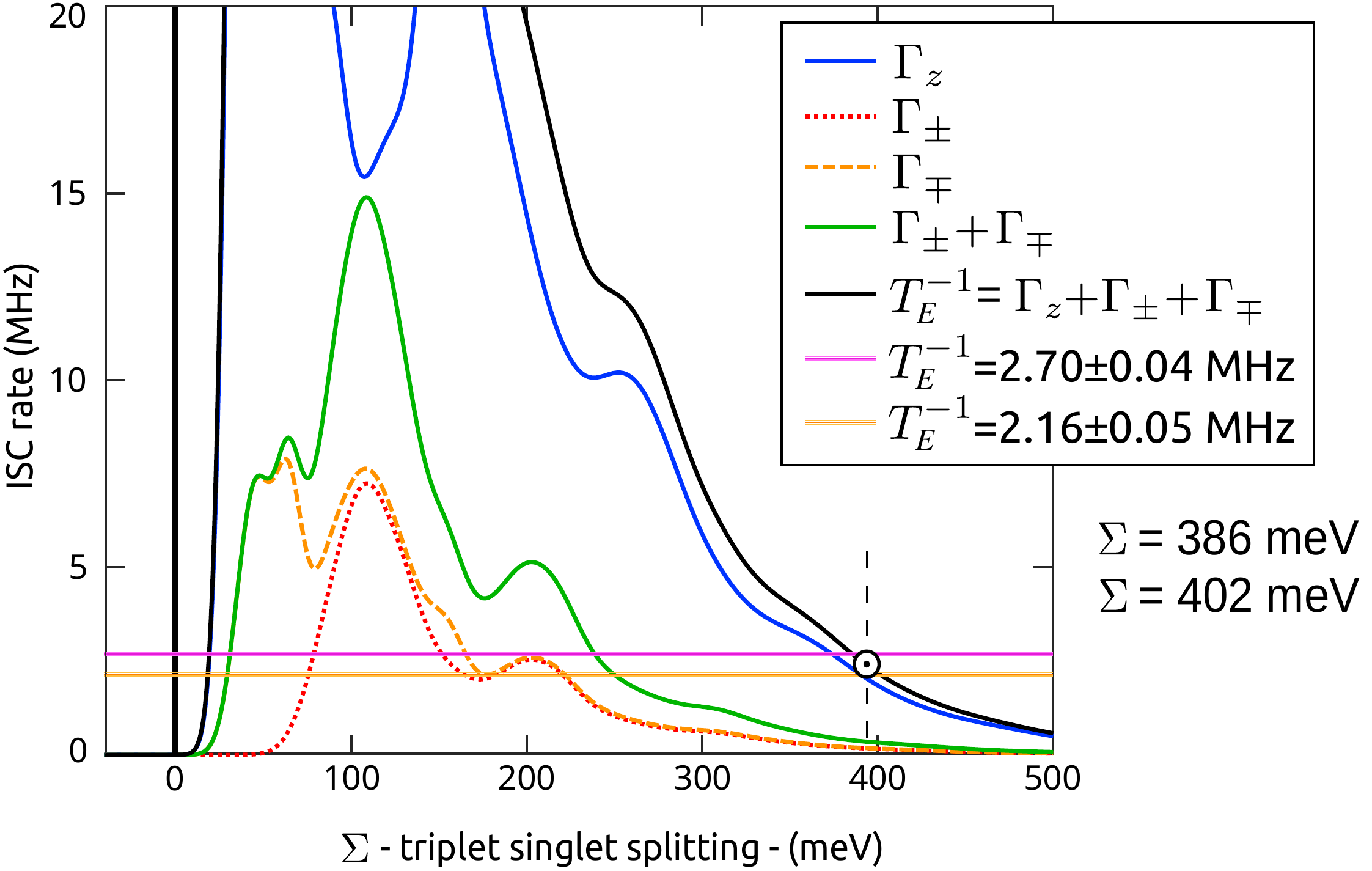} 
	\caption{Calculated low-temperature ISC rates ($\Gamma_{z}$, $\Gamma_{\pm}$, $\Gamma_{\mp}$ from Eqs.~\eqref{eq:gamma1}-\eqref{eq:gamma3} as a function of the energy gap ($\Sigma$) between the shelving state singlet state and the triplet ground state. Here we applied $\lambda_\perp=1.2\lambda_z$ by following Ref.~\onlinecite{Goldman2015a}. Two experimental data about the lifetime of the singlet state ($T_E$) is applied from Ref.~\onlinecite{Acosta2010} with $T_E^{-1}\approx$2.16~MHz (yellow horizontal line) and Ref.~\onlinecite{Robledo2011} with $T_E^{-1}\approx$2.70~MHz (purple horizontal line). The crossing point between the simulated inverse lifetime (black curve) and the experimental ones is depicted by a circle that results in $\Sigma$=402~meV and $\Sigma$=386~meV, respectively.} 
	\label{fig:SOCtrend}
\end{figure}

\begin{figure}[ht] 
	\includegraphics[width=7.8cm]{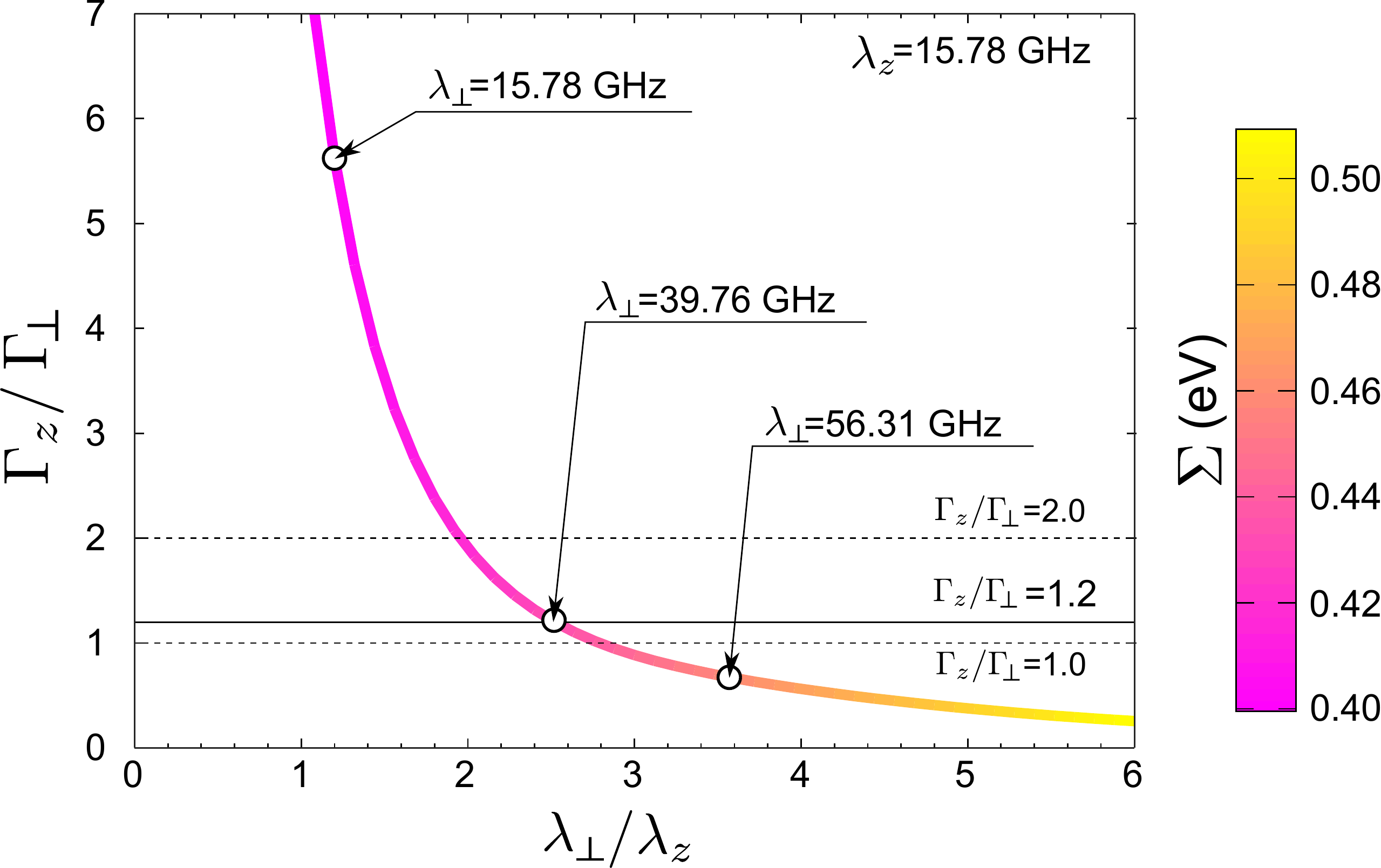} 
	\caption{$\Gamma_z/\Gamma_\perp$ is plotted as a function of $\lambda_\perp/\lambda_z$ where $\lambda_z$=15.78~GHz is our accurate DFT value.  The $\lambda_\perp$=56.31~GHz value approximated from DFT wavefunctions  is an overestimation. $\lambda_\perp$=39.76~GHz yields $\Gamma_z/\Gamma_\perp$=1.2.} 
	\label{fig:G_vs_lambda}
\end{figure}

\begin{figure}[ht] 
	\includegraphics[width=6.8cm]{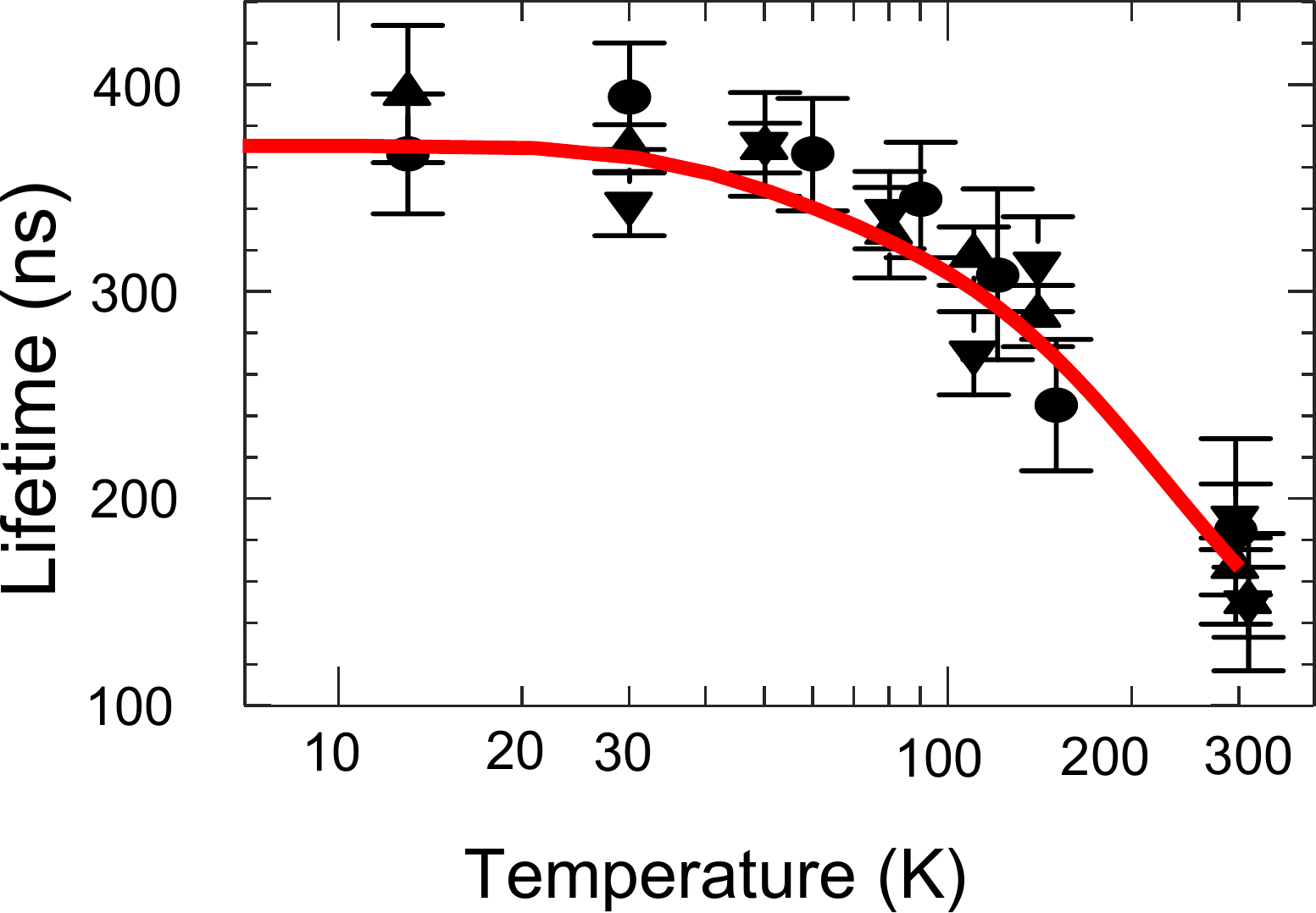} 
	\caption{The calculated lifetime of the singlet shelving state is plotted as a function of the temperature with the observed lifetimes for two single NV centers (dot and triangle data points with uncertainties) taken from Ref.~\onlinecite{Robledo2011}.} 
	\label{fig:lifetime_T}
\end{figure}

We previously calculated all the parameters from \emph{ab initio} wavefunctions required to calculate the ISC rates that are plotted in  Fig.~\ref{fig:SOCtrend} and compared to the observed inverse lifetime of the singlet~\cite{Acosta2010, Robledo2011}. 
We find that the energy gap between the shelving singlet state and the triplet ground state is $\approx$0.4~eV. This is very reasonable as the sum of the previously deduced $\Delta$ (Refs.~\onlinecite{Goldman2015a, Thiering2017}), the observed $\Lambda$ (Ref.~\onlinecite{Rogers2008}) and our deduced $\Sigma$ approximately equals to the ZPL energy between the triplets [c.f., Fig.~\ref{fig:NV}(c)]. With the present choice of the $\lambda_\perp=1.2~\lambda_z$, $\Gamma_z$ ISC rate (blue curve) is about $6\times$ larger than the $\Gamma_\perp = \Gamma_\pm + \Gamma_\mp$ rate (green curve) at that energy gap that would further strengthen the spinpolarization process beside the strictly spin-selective process in the upper branch. On the other hand, Robledo and co-workers deduced a smaller $\Gamma_z / \Gamma_\perp\approx 1.1\dots2$ from experimental data on two single NV centers at room temperature~\cite{Robledo2011} where the common value was 1.20 by taking the uncertainty in the measurements into account. By using the low-temperature simulation data, we varied the $\lambda_z / \lambda_\perp$ and plotted $\Gamma_z / \Gamma_\perp$ in Fig.~\ref{fig:G_vs_lambda} to analyze this issue, as there is uncertainty in the value of $\lambda_\perp$.     
We conclude that $\lambda_\perp/\lambda_z\approx2$ is required to obtain the experimentally deduced ratio between the ISC rates. Since the experimental values for the ratio between the ISC rates scattered about 100\% in the two individual NV centers we conclude that the accurate ratio of ISC rates should be further investigated in the experiments, and we do not rely on these experimentally deduced data. We rather used the $\lambda_\perp=1.2 \lambda_z$ in the study of the temperature dependence of the ISC rates that we plot in Fig.~\ref{fig:lifetime_T} (red curve) and compare to previous experimental data taken on two single NV centers.    
Here we used the calculated vibronic states of the $|{}^{1}\widetilde{E}\rangle$ with the Boltzmann occupation of vibronic levels at the given temperature, in order to compute ISC rates as defined in Eqs.~\eqref{eq:gamma1}-\eqref{eq:gamma3}. We found a very good agreement with the experimental data~\cite{Robledo2011} as the calculated lifetime is reduced from 370~ns at cryogenic temperatures down to 171~ns at room temperature to be compared to 371$\pm$6~ns and 165$\pm$10~ns, respectively. Our calculations reveal that the vibronic state associated with the optically forbidden phonon feature at $\approx$14~meV in the PL spectrum plays a key role in the temperature dependence of the ISC rates. The calculated $\Gamma_z/\Gamma_\perp$ is only reduced by $\sim$5\% going from cryogenic temperature to room temperature which means that the spinpolarization efficiency per single optical cycle does not degrade significantly. These results demonstrate that our theory can account for the intricate details of the ISC processes in NV center and reproduce the basic experimental data.

\section{Summary and conclusions}
\label{sec:6}

In this work, we developed a theory on the nature of the singlet states including electron-electron correlation coupled with phonons. We identified the strong electron phonon coupling between the singlet states that can be described as a combination of pseudo Jahn-Teller effect and damped dynamic Jahn-Teller effect. We extended the theory of ISC rates of NV center to account for this complex nature of the singlet shelving state that is responsible for the ISC process toward the ground state. Our theory can explain several features in the optical spectra of singlets. In particular, the presence of optically forbidden state in unstressed diamond and the features in the phonon sideband of the PL spectrum was well reproduced which are based on the vibronic nature of the singlet shelving state. The calculated ISC rates and the deduced energy gap between the shelving singlet state's level and the triplet ground state's level are consistent with the previous experimental data. The calculated temperature dependence of the lifetime of the singlet shelving state is in very good agreement with the experimental data. Our results complete the theoretical description of the entire optical spinpolarization loop of NV center. 

Our results may have an impact in the field, as NV center is a template for similar defects that act as solid state qubits. The most obvious example is the neutral divacancy in silicon carbide~\cite{Gali2011, Koehl2011} for which the first step for understanding the underlying mechanisms has been recently taken~\cite{Christle2017}. Our \emph{ab initio} toolkit can be extended to other defect systems including point defects in 2D materials such as boron nitride, transition metal dichalgonides and dioxides that have been attracted a great attention. Computing the ISC rates of these defects can contribute to understand their optical properties and optimize their quantum bit operation.
 
 \section*{Acknowledgment}
Support from the Hungarian Government and the National Research 
Development and Innovation Office (NKFIH) in the frame of the \'UNKP-17-3-III 
New National Excellence Program of the Ministry of Human Capacities, NVKP 
Project Grant No.\ NVKP\_16-1-2016-0043 and the Quantum Technology 
National Excellence Program (Project No.\ 2017-1.2.1-NKP-2017-00001) are 
acknowledged. 

\appendix
\section{Pseudo Jahn-Teller effect from perturbation theory}
\label{app:A}

We derive the PJT effect between $^1E$ and $^1A_1$ from perturbation theory that provides an insight about the strength of interaction where we concentrate on the vibronic ground state of the resulting $\left|^{1}\widetilde{E}\right\rangle$. The vibronic wavefunction from $^1A_1$ coupled to the $^1E$ should transform as $E$, thus the $E$ phonon state should be occupied in $^1A_1$ state with an effective $\hbar \omega_E$ phonon energy. The energy difference between the corresponding states will be $\Lambda_e + \hbar \omega_E$ where $\Lambda_e$ is the energy gap between the electronic levels of $^1E$ and $^1A_1$. Now by choosing the $x$ representation from the double degenerate states and labeling the $|E_xE_y\rangle$ vibration wavefunction by the occupation representation we arrive at  
\begin{equation}
\left|^{1}\widetilde{E}_{x}\right\rangle =\left|^{1}E_{x}\right\rangle \otimes\left|00\right\rangle +\frac{\chi_{\mathrm{PJT}}}{\Lambda+\hbar\omega_{E}}\left|^{1}A_{1}\right\rangle \otimes a_{x}^{\dagger}\left|00\right\rangle 
\label{eq:AppA1}
\end{equation}
with the $\chi_\mathrm{PJT}$ is the coupling parameter and $ a_{x}^{\dagger}$ is the creation operator of the $E_x$ phonon. In the Kohn-Sham DFT calculations $\Lambda_e$=0 for $|xx\rangle$ singlet state resulting in relatively large Jahn-Teller energy, however, it can be seen in Eq.~\eqref{eq:AppA1} that the strength of the interaction is significantly damped by 
$\Lambda_e\approx$1.19~eV for the true singlet eigenstates. The coupling parameter can be calculated as
\begin{equation}
\chi_\mathrm{PJT}=\left\langle ^{1}E_{x}\right|\otimes\left\langle 00\right|\hat{H}_\mathrm{PJT}\left|^{1}A_{1}\right\rangle \otimes a_{x}^{\dagger}\left|00\right\rangle \text{,}
\label{eq:AppA2}
\end{equation}
where $\hat{H}_{\mathrm{PJT}}$ is the PJT Hamiltonian from Eq.~\eqref{eq:PJT4}.
By substituting the $\hat{H}_\mathrm{PJT}$ into Eq.~\eqref{eq:AppA2}, we arrive at 
\begin{equation}
\chi_\mathrm{PJT}=\frac{\tilde{F}}{\sqrt{2}}\left\langle ^{1}E_{x}\right|\sigma_{z}\left|^{1}A_{1}\right\rangle
\label{eq:AppA4}
\end{equation}
where we used $\left\langle 00\right|a_{y}a_{y}^{\dagger}\left|00\right\rangle$=0 and $\left\langle 00\right|a_{x}a_{x}^{\dagger}\left|00\right\rangle$=1 relations. As a next step we use the two-particle expression of $^{1}A_{1}$ in Eq.\ref{eq:PJT2} to arrive at
\begin{equation}
\chi_\mathrm{PJT}=\frac{\tilde{F}}{\sqrt{2}}\left[\left\langle xx\right|-\left\langle yy\right|\right]\sigma_{z}\left[\left|xx\right\rangle +\left|yy\right\rangle \right]
\label{eq:AppA5}
\end{equation}
and finally solve it in the matrix representation as 
\begin{equation}
\chi_\mathrm{PJT}=\frac{\tilde{F}}{\sqrt{2}}\frac{1}{2}\begin{pmatrix}1 & 0  & -1\end{pmatrix}\begin{pmatrix}1\\
& 0\\
&  & -1
\end{pmatrix}\begin{pmatrix}1\\ 0
\\
1
\end{pmatrix}=\frac{\tilde{F}}{\sqrt{2}} \text{.}
\label{eq:AppA6}
\end{equation}
This completes the perturbation theory on the PJT effect, in order to give insight about the nature of this interaction. We note that we obtained our results by direct diagonalization of the full electron-phonon Hamiltonian in the main part of the paper which goes beyond this perturbation theory.	
		
\section{Vibronic wavefunctions \label{app:vibwfs}}
	
Here we show the calculated coefficients of the $^1\widetilde{E}\oplus^1\widetilde{A}_1$ vibronic wavefunctions in Table~\ref{tab:coeffs}. 	
\begin{table*}[htbp]
 \caption{Coefficients are defined in Eqs.~\eqref{eq:Epolaron} and \eqref{eq:Epolaron2} for the $^1\widetilde{E}$ and $^1\widetilde{A}_1$ vibronic states, respectively. The first column defines the phonon index ($n$). The
"repr. of phonons" column shows the irreducible representation of states that can be constructed from $n$ phonons. The $A_2$ modes are negligible and labeled in the parentheses. We determined the coefficients up to $n=10$ phonon limit, but we present the rows only up to $n=6$ since all of the $n>6$ coefficients are below 0.001. We refer to the summation of all the individual coefficients with $n$ phonons by $\underset{{\textstyle n_{i}=n}}{\sum}$. The ground vibronic state of $^1\widetilde{E}$ that transforms as $E$ is expressed by $c_1^2$, $d_1^2$, and $f_1^2$. The first excited vibronic wavefunction of $^1\widetilde{E}$ that transforms as $A_1$ is expressed by $c^{\prime2}_1$ and  $d^{\prime2}_1$.}
 \begin{ruledtabular}
 \begin{tabular}{ccccccc}
 & \multicolumn{3}{c}{$^1\widetilde{E}$} & \multicolumn{2}{c}{$^1\widetilde{A}_1$} & \\
$n$ & $\underset{n_{i}=n}{\sum}c_{i}^{2}$ & $\underset{n_{i}=n}{\sum}d_{i}^{2}$ & $\underset{n_{i}=n}{\sum}f_{i}^{2}$  & $\underset{n_{i}=n}{\sum}c_{i}^{\prime2}$ & $\underset{n_{i}=n}{\sum}d_{i}^{\prime2}$& repr. of phonons \tabularnewline
\vline
0 & $c_{1}^{2}=$0.645 & - & -  & $c_{1}^{\prime2}=$0.017  & -& $A_{1}$\tabularnewline
1 & - & $d_{1}^{2}=$0.029 & $f_{1}^{2}=$0.063  & - & $d_{1}^{\prime2}=$0.618& $E$\tabularnewline
2 & $c_{2}^{2}=$0.090 & $d_{2}^{2}=$0.004 & $f_{2}^{2}=$0.089  & $c_{2}^{\prime2}=$0.045 & $d_{2}^{\prime2}=$0.042& $A_{1}+E$\tabularnewline
3 & $c_{3}^{2}=$0.011 & $d_{3}^{2}=$0.012 & $f_{3}^{2}=$0.012  & $c_{3}^{\prime2}=$0.004  & $d_{3}^{\prime2}=$0.194& $A_{1}+(A_{2})+E$\tabularnewline
4 & $c_{4}^{2}=$0.015 & $d_{4}^{2}+d_{5}^{2}=$0.002 & $f_{4}^{2}+f_{5}^{2}=$0.016  & $c_{4}^{\prime2}=$0.016 & $d_{4}^{\prime2}+d_{5}^{\prime2}=$0.018& $A_{1}+2E$\tabularnewline
5 & $c_{5}^{2}=$0.002 & $d_{6}^{2}+d_{7}^{2}=$0.003 & $f_{6}^{2}+f_{7}^{2}=$0.002  & $c_{5}^{\prime2}=$0.002 & $d_{6}^{\prime2}+d_{7}^{\prime2}=$0.032& $A_{1}+(A_{2})+2E$\tabularnewline
6 & $c_{6}^{2}+c_{7}^{2}=$0.002 & $d_{8}^{2}+d_{9}^{2}=$0.000 & $f_{8}^{2}+f_{9}^{2}=$0.002  & $c_{6}^{\prime2}+c_{7}^{\prime2}=$0.003  & $d_{8}^{\prime2}+d_{9}^{\prime2}=$0.003& $2A_{1}+(A_{2})+2E$\tabularnewline
... & ... & ... & ... & ... & ... & ...\tabularnewline
 \end{tabular}
 \end{ruledtabular}     
 \label{tab:coeffs}
\end{table*}

\section{Transition dipole moment between the singlet states\label{App:dipole}}

Here we determine the optical transition strengths between
$\left|^{1}A_{1}\right\rangle$ and $\left|^{1}E\right\rangle$. Following the derivation of Hepp \textit{et al.} in Eq.\ (13) in the Supplemental Material of Ref.~\onlinecite{Hepp2014}, the transition dipole moments
between the single particle orbitals in $C_{3v}$ symmetry with polarization "$x$" are the followings
\begin{equation}
\begin{array}{cc}
"x" & \begin{array}{cc}
\!\!\bigl|e_{x}\bigr\rangle & \!\!\bigl|e_{y}\bigr\rangle\end{array}\\
\begin{array}{c}
\bigl\langle e_{x}\bigr|\\
\bigl\langle e_{y}\bigr|
\end{array} & \!\!\begin{bmatrix}d_{\perp}\\
& -d_{\perp}
\end{bmatrix}
\end{array}\qquad\begin{array}{cc}
"y" & \begin{array}{cc}
\bigl|e_{x}\bigr\rangle & \bigl|e_{y}\bigr\rangle\end{array}\\
\begin{array}{c}
\bigl\langle e_{x}\bigr|\\
\bigl\langle e_{y}\bigr|
\end{array}\!\! & \begin{bmatrix} & -d_{\perp}\\
-d_{\perp}
\end{bmatrix}
\end{array} \text{.}
\label{eq:Lum1}
\end{equation}
By using the relations in Eq. \eqref{eq:Lum1}, the dipole moment for the two-particle wavefunctions can be expressed and applied to get 
\begin{equation}
P\left(^{1}A_{1}\overset{"x"}{\leftrightarrow}{}^{1}E_{x}\right)=\Bigl|\underset{A_{1}}{\underbrace{\left\langle ^{1}A_{1}\right|}}\underset{E}{\underbrace{\hat{d}_{x}^{(1)}+\hat{d}_{x}^{(2)}}}\underset{E}{\underbrace{\left|^{1}E_{x}\right\rangle }}\Bigr|^{2}=4d_{\perp}^{2}
\label{eq:Lum2}
\end{equation}
where we introduced the transition optical dipole operator ($\hat{d}_{x}^{(1)}$ and $\hat{d}_{x}^{(2)}$) acting on particle 1 and 2, respectively. Our result is in agreement with a previous result (see Tab.\ A.4 in Ref.~\onlinecite{Doherty2011}, where the $2 O_{b,x}$ matrix element is the transition dipole moment, thus the transition strength is $4 O^2_{b,x}$ that is corresponding to our $4d_{\perp}^{2}$ in Eq.~\eqref{eq:Lum2}). The three other possible transitions are
\begin{equation}
P\left(^{1}A_{1}\overset{"y"}{\leftrightarrow}{}^{1}E_{y}\right)=4d_{\perp}^{2}
\label{eq:Lum3}
\end{equation}
and
\begin{equation}
P\left(^{1}A_{1}\overset{"x"}{\leftrightarrow}{}^{1}E_{y}\right)=P\left(^{1}A_{1}\overset{"y"}{\leftrightarrow}{}^{1}E_{x}\right)=0 \text{.}
\label{eq:Lum4}
\end{equation}
The dipole operator can be expressed in a similar form as the PJT Hamiltonian [see Eq. \eqref{eq:PJT4}],
\begin{equation}
\hat{d}_{x}=2d_{\perp}\hat{\sigma}_{z}\qquad\qquad\hat{d}_{y}=2d_{\perp}\hat{\sigma}_{x} \text{,}
\label{eq:Lum6}
\end{equation}
where $\hat{\sigma}_{z,x}$ matrices are defined in Eq.~\eqref{eq:PJT5}. As an example, we explicitly write the intensity of the ZPL transition between the vibronically coupled singlet states as follows 
\begin{equation}
I_{\mathrm{ZPL}}=\left|\left\langle ^{1}\widetilde{E}_{x}\right|\hat{d}_x\left|^{1}\widetilde{A}_{1}\right\rangle \right|^{2}=4d_{\perp}^{2}\left(\sum_{i=1}^{\infty}c_{i}c_{i}^{\prime}+\sum_{i=1}^{\infty}d_{i}d_{i}^{\prime}\right)
\label{eq:Lum7}
\end{equation}
where the corresponding expansion coefficients ($c_i, c_i^\prime, d_i, d_i^\prime$) are defined in Eqs.~\eqref{eq:Epolaron} and \eqref{eq:Epolaron2}. 
\label{app:B}



%

\end{document}